\documentclass[12pt]{article}
\usepackage{graphicx,color,url}
\usepackage{amsmath}
\usepackage{subfigure}
\usepackage{setspace} 
\usepackage{lineno}  
\usepackage{authblk} 
\usepackage{array}
\usepackage{caption}
\usepackage{booktabs}
\usepackage{amssymb}
\usepackage[authoryear]{natbib}
{}
\renewcommand{\cite}[1]{[\citeauthor{#1}, \citeyear{#1}]}
\graphicspath{ {../images/} }
\newcommand{\figwidth}{0.5\textwidth}

\usepackage{helvet}

\usepackage[top=2.54cm, bottom=2.54cm, left=2.54cm, right=2.54cm]{geometry}
\newcolumntype{R}[1]{>{\raggedright\arraybackslash}p{#1}}
\begin{document}

\title{Energetic closure of the\\spatially resolved global food system}
\author[1,3]{Maxwell Kaye}
\author[2]{Graham K. MacDonald}
\author[3]{William Fajzel}
\author[3,4]{Eric Galbraith}
\affil[1]{Department of Mathematics, McGill University}
\affil[2]{Department of Geography, McGill University}
\affil[3]{Department of Earth and Planetary Sciences, McGill University}
\affil[4]{ICREA, ICTA-UAB}
\date{}
\maketitle
\doublespacing{}

\abstract{Integrated global food system analysis is hampered by the fragmentation of data among food types, processes, and scales. Studies also often neglect the connection to human metabolism—the ultimate driver of food demand. Here we use a common energetic framework to harmonize data on 95 individual food commodities across food system processes, including production, processing, animal feed and consumption, and estimate human metabolism from body size, demographic, and activity data. We estimate that the share of unmetabolized food calories globally doubled between 1990 and 2019 (from about 10 to 20\% of the total calories available for human consumption) as food supply outpaced energy expenditure. Approximately half (51\%) of the global population’s metabolic demands could theoretically be met by production in the same local 1-degree grid cell (~10,000 km$^2$) when holding diets constant. Our open-source framework can be applied to assess strategies to reduce food system inefficiencies from photosynthesis to metabolism while meeting local energetic demands.}

\clearpage
Food energy is conserved from initial production to final utilization. However, the availability and heterogeneity of food systems data are substantial impediments to holistically capturing food flows globally \cite{Kim_2021}. Inconsistencies in census data across countries and food types (including crops, livestock and fisheries) can also make it difficult to compare data between sources \cite{Enthoven_Van_den_Broeck_2021, kebedeAssessingAddressingGlobal2024}, revealing a need for systematic data harmonization approaches. \cite{Xue_2017} stressed the need for consistent, comparable, and publicly available databases, which would allow flexible analysis of entire food supply chains \cite{Kummu_2012, Vittuari_2019, Karakoc_Konar_Puma_Varshney_2023, zhaoTraceableScalableFood2025}. In relation to this need, the OECD recently identified inconsistencies across methodologies and lack of detail as significant inhibitors to progress in food systems research \cite{Deconinck_2021}. Challenges in collating food systems data sources arise from differences in the resolution and coverage across three dimensions: scale (spatial and temporal extent, including subnational detail), commodity (number of individual crop, livestock, and fisheries products and degree of commodity aggregation), and processes (production, transformation, and final utilization and metabolism).

Advances in remote sensing and other computational approaches have supported increased availability of global gridded food production data ~\cite{Herrero_2017, GAEZ_2022, MAGPIE_Dietrich_2019, tubiello_2023, SPAM_2019, davisHarvestStatGlobalEffort2025, Monfreda_Ramankutty_Foley_2008}. While such data paint a detailed picture of agricultural land use for food production, the connection to other processes including final human metabolic demand remains more obscure. Global diets have separately been assessed in the context of sustainability and public health \cite{herforthHealthyDietBasket2025, Willett_2019}; however, connecting diets to original production in a spatially resolved manner is a considerable challenge \cite{Cassidy_2013}, often involving study- or model-specific calculation steps and assumptions. The gridded crop production datasets underlying these analyses also vary substantially in terms of the spatial patterns of crops, often with infrequent but major updates \cite{tangCROPGRIDSGlobalGeoreferenced2024, Tubiello_Conchedda_2023, SPAM_2019}. Aquatic and terrestrial food data are often separated entirely or presented in a highly aggregated form with low commodity resolution \cite{wasseniusGlobalAnalysisPotential2023}. While recent work has provided a national scale assessment of food production including both agriculture and aquaculture distinguished by food commodity \cite{wasseniusGlobalAnalysisPotential2023}, production data remains disjointed from the human metabolism that fundamentally drives food demand.

Energetic units (kilocalories or Joules) provide a logical bridge between Earth system, environment, and nutrition/public health aspects of the food system. Prior research has developed energetic approaches to specific aspects of the food system \cite{Kastner_2021} including calculating the calories of crop production which support human diets \cite{Cassidy_2013}, the spatial distribution of embodied crop calories in animal products \cite{pradhanEmbodiedCropCalories2013}, human appropriation of net primary production (HANPP) \cite{Haberl_Erb_2007}, as well as Food Loss and Waste (FLW) \cite{alexander_2017, Kummu_2012}. While previous global studies have considered gridded food production in relation to diets \cite{Herrero_2017, DeFries_2015, Kinnunen_2020, wasseniusGlobalAnalysisPotential2023}, to our knowledge, these studies have not captured energy flows from production to metabolism while maintaining a high commodity resolution and coverage.

Global food system data can be used to assess food self-sufficiency, the extent to which regions (usually countries or other political jurisdictions \cite{ImplicationsEconomicPolicy}) could theoretically sustain their populations without imports \cite{clappFoodSelfsufficiencyMaking2017}. Definitions include the ratio of “dietary energy demand” to “dietary energy supply” \cite{porkkaFoodInsufficiencyTrade2013, baer-nawrockaFoodSecurityFood2019, beltran-penaGlobalFoodSelfsufficiency2020}, measures of the number of nutrients available in the food supply ~\cite{wasseniusGlobalAnalysisPotential2023}, whether a threshold condition is met in each region \cite{pradhanFoodSelfSufficiencyScales2014}), or the number of people that could be sustained within a transport-optimized distance between production and demand \cite{Kinnunen_2020}. Prior food self-sufficiency estimates have focused on various spatial and commodity scales, including aggregated food production and across key food groups   \cite{porkkaFoodInsufficiencyTrade2013,pradhanFoodSelfSufficiencyScales2014,wasseniusGlobalAnalysisPotential2023,Kinnunen_2020}. While estimated national food supply (food available for consumption by humans) is often used as a proxy for diet in the food loss and waste (FLW) and food self-sufficiency (FSS) literature, it remains unclear how actual human metabolic demand (the energy required to sustain humans) compares with production and food supplies, especially when considering many individual food commodities and disparate food groups. 

We present and analyze a unified, gridded dataset that links commodity-specific food production—from crops, livestock, and marine fisheries—to human metabolism, using energy as a common conserved quantity that flows across space, food groups, and stages of the food system. We focus on 2015 as a representative year but also analyze trends over time. We estimate metabolic rates from country-specific data and scale it to grid cells based on population. These rates are then compared with crop production, feed, animal production, and food supply to close the loop from production to bioenergetics, and to assess self-sufficiency. Grounded in the principle that all food must be metabolized, otherwise used, or lost, our approach provides a new perspective on systemic inefficiencies in the food system \cite{alexander_2017} and supports strategies toward integrated human–Earth system modelling \cite{liuSystemsIntegrationGlobal2015}. All data and Python code are openly available.
\section*{Results}
\subsection*{Global food balance}

\begin{figure}[htbp]
    \centering
    \includegraphics[height=550pt]{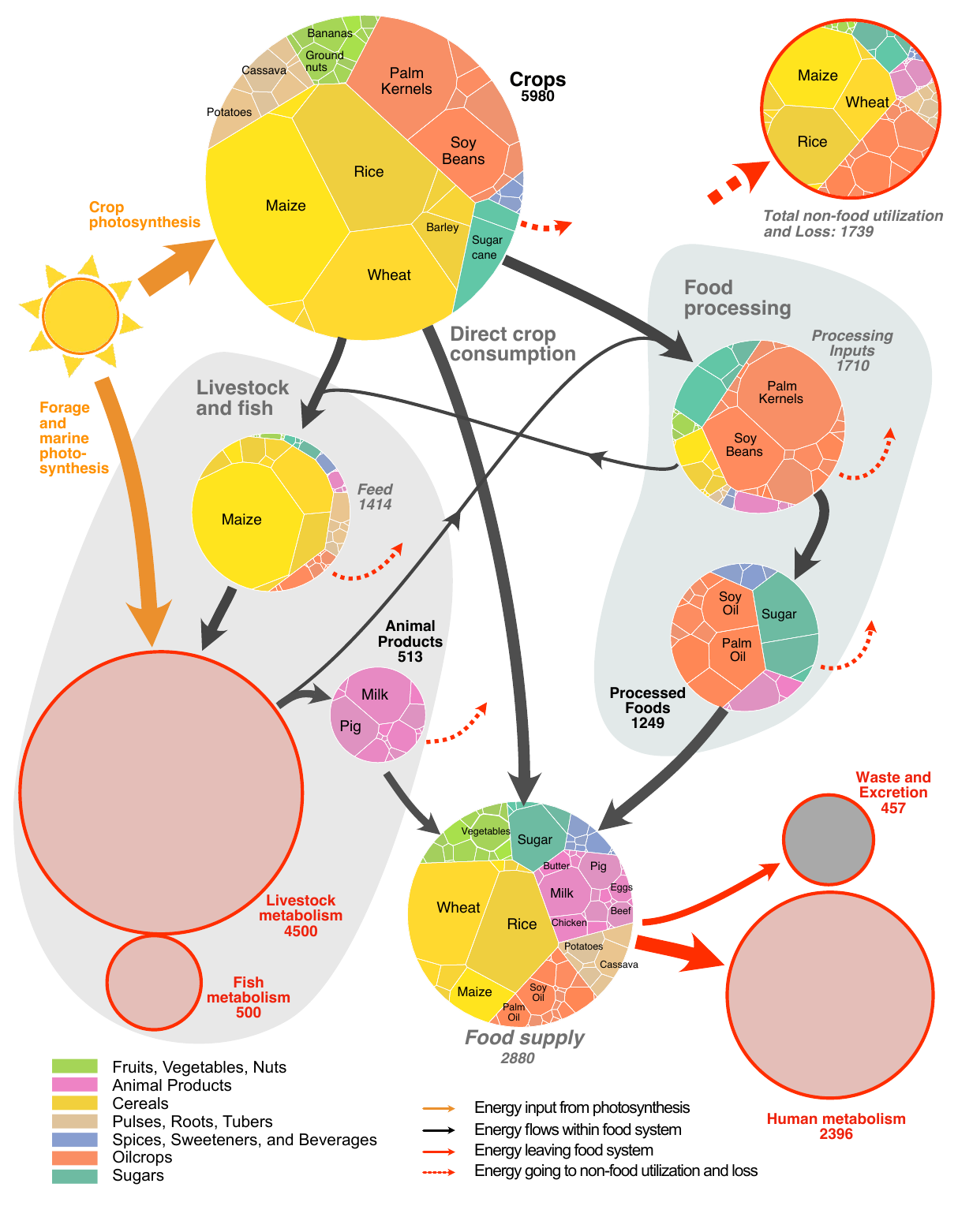}
    \caption{Energy flows in the global food system. Voronoi Tree-maps show the composition at each stage of the food system, with areas proportional to the corresponding energetic flows. All food energy, ultimately sourced from photosynthesis, is accounted for until its final fate as metabolic energy, loss, or waste. All quantities are given in units of global kcal per human per year. The sum of production terms (Crops, Animal Products, and Processed Foods) shown in bold is equal to the sum of terms shown in gray italics. Metabolism estimates reflect the energy conversion rate of respiration by humans, all livestock and the wild and aquaculture fish that contribute to the food supply. See the supplementary information for fully-labeled tree-maps.}\label{fig:energyFlowVornoi}
\end{figure}

We first estimate nationally aggregated energetic food flows by food commodity, production type, utilization category, and metabolism for humans and food-producing animals. Using a quadratic optimization algorithm to minimize relative change, these values are adjusted with trade data to satisfy the law of conservation of energy. Figure ~\ref{fig:energyFlowVornoi} depicts our globally aggregated accounting with the total flows of each food group through the food system. Our estimate of global per-capita human metabolism (2396 kcal/cap/day) is 40\% of the total annual crop production energy (5980 kcal/cap/day).

Food energy from crop production sources (Figure \ref{fig:energyFlowVornoi}) primarily flows to one of three sinks—feed, processing, or directly to food supply. The feed and processing sinks then transform the energy into new food commodities before entering the food supply (either via animal production or food processing). We estimate that, in 2015, approximately 29\% of calories were lost from the food system or used for other non-food uses (e.g., bioenergy and other industrial use). This agrees well with prior estimates of energetic food losses, which range from approximately 24\% \cite{Kummu_2012} to 36\% \cite{alexander_2017} circa the year 2011, while our human food consumption and crop production values are slightly higher. The annual food supply represents about 48\% of the annually produced crop energy, and 83\% of the food supply was metabolized by humans.
\subsection*{Spatial patterns}
\begin{figure}[htbp]
    \centering
    \begin{minipage}[b]{0.6\textwidth}
        \raggedright (a) Food Production\\
        \includegraphics[width=\linewidth,clip,trim=0 2.5cm 0 0]{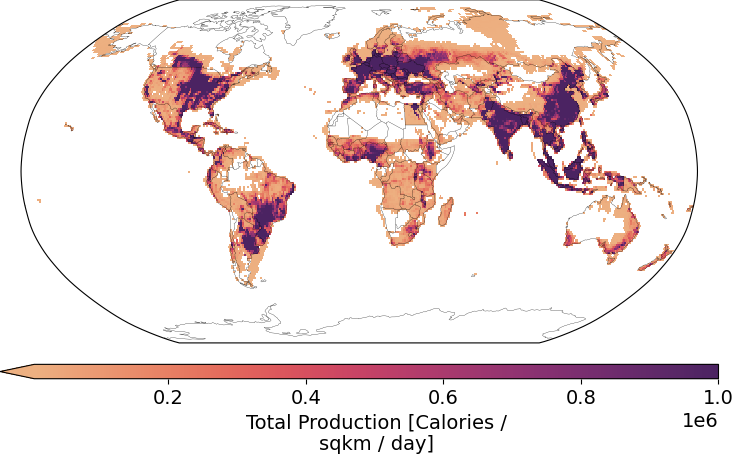}
    \end{minipage}
    \vspace{0.1cm}
    
    \begin{minipage}[b]{0.6\textwidth}
        \raggedright (b) Human Metabolism\\
        \includegraphics[width=\linewidth,clip,trim=0 1.1cm 0 0]{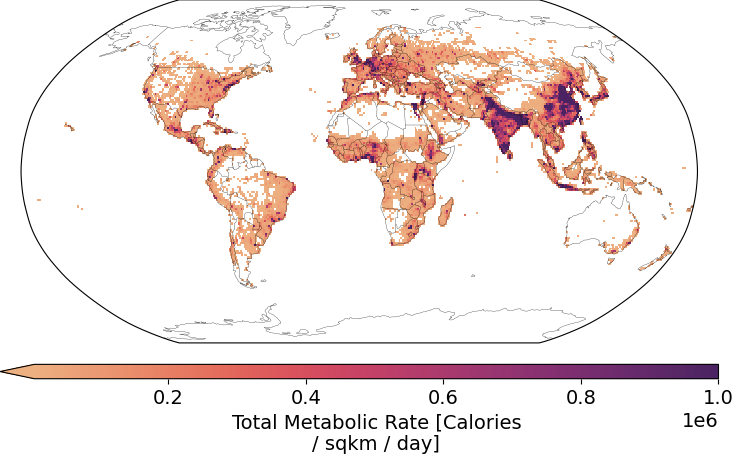}
        \centering \tiny Million kcal/km$^2$/day
    \end{minipage}
    \vspace{0.1cm}
    
    \begin{minipage}[b]{0.6\textwidth}
        \raggedright (c)\\
        \includegraphics[width=\linewidth, clip, trim=0 0.5cm 0 0.0]{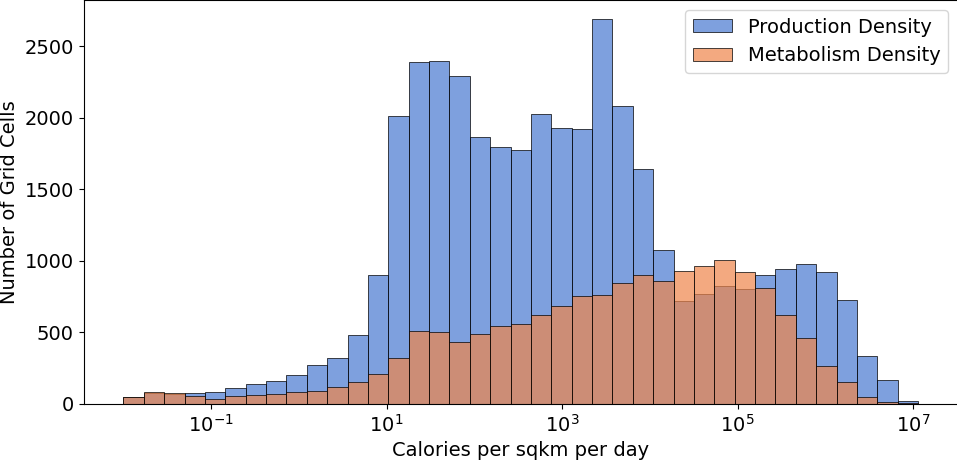}
        \centering \tiny kcal/km$^2$/day
    \end{minipage}
    \caption{Spatial distribution of total food production and human metabolized calories, summed within each grid cell, including crops and animal products. (a) Total production of food balance sheet items. In order to avoid double counting, non-primary food production rates (processed foods) are not included. For lack of surrogate data, freshwater fish and aquaculture are not shown ($<1\%$ of total). (b) Total human metabolism per square meter, obtained by distributing calculated national average metabolic rates onto population counts. A cut-off of 5000 kilocalories/square kilometer/day was used to mask grid cells with negligible production in the map. (c) Histogram of densities.}\label{fig:prod_cons}
\end{figure}

Geographically, food production and consumption energy densities range from negligible to millions of kilocalories per km$^2$ per day (Figure \ref{fig:prod_cons} a \& b). While food energy production often tends to be concentrated nearby consumption, there are important differences. The unimodal consumption distribution has left-skew on logarithmic axes~(Figure \ref{fig:prod_cons}c), intuitively reflecting that most metabolism occurs in population-dense locations (especially urban areas). In contrast, the production distribution is multimodal, reflecting the contrast of typically more intensive production of some crops around major population centers \cite{theboGlobalAssessmentUrban2014} as well as more remote extensive grain production, livestock grazing, and fisheries. This distributional asymmetry between production and consumption is an important qualitative aspect of the spatially-resolved global food system, requiring that food flows from broad regions of production to more focused sites of consumption.

\begin{figure}[htbp]
    \centering
    \begin{tabular}{c c}
        \begin{minipage}[b]{\figwidth}
            \centering
            \raggedright (a)\\
            \includegraphics[width=\linewidth]{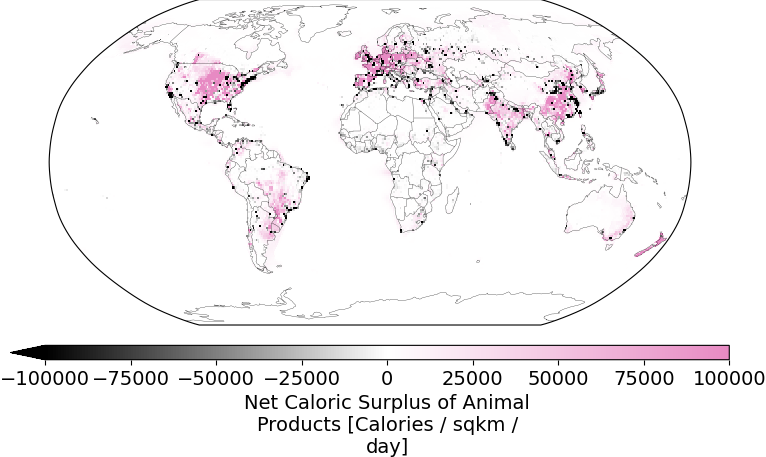}
        \end{minipage} &
        \begin{minipage}[b]{\figwidth}
            \centering
            \raggedright (b)\\
            \includegraphics[width=\linewidth]{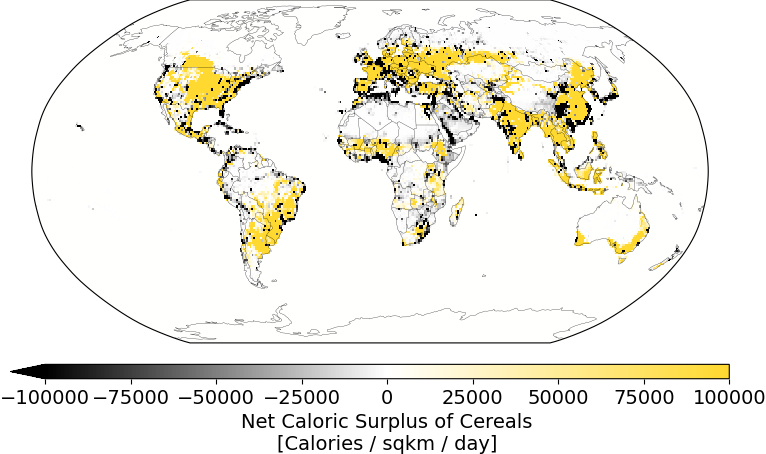}
        \end{minipage} \\
        \begin{minipage}[b]{\figwidth}
            \centering
            \raggedright (c)\\
            \includegraphics[width=\linewidth]{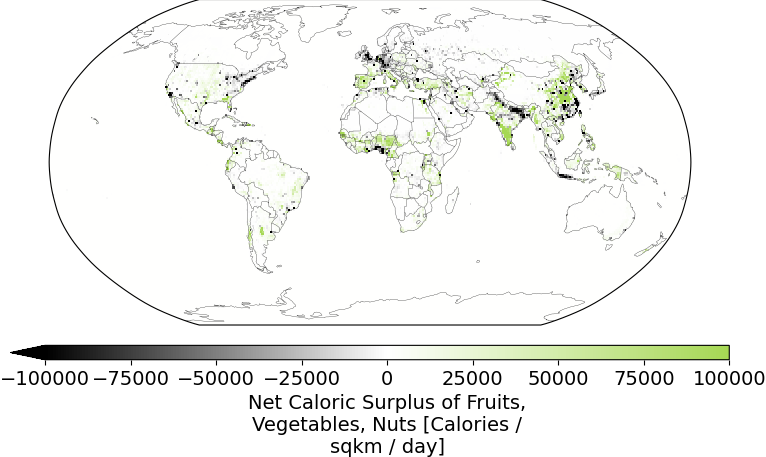}
        \end{minipage} &
        \begin{minipage}[b]{\figwidth}
            \centering
            \raggedright (d)\\
            \includegraphics[width=\linewidth]{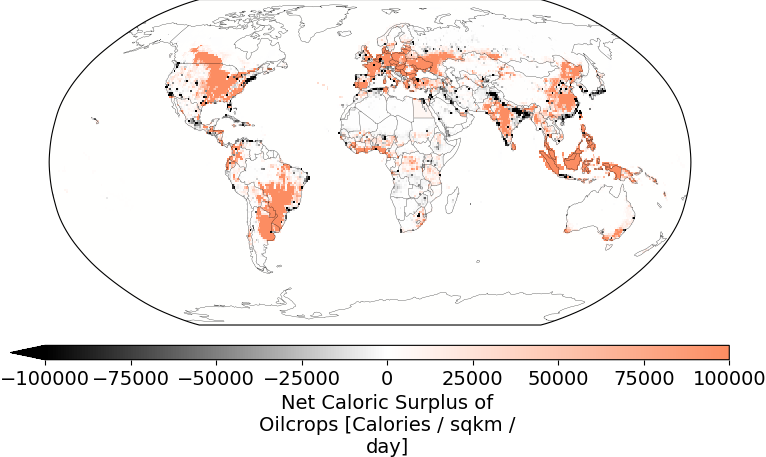}
        \end{minipage} \\
        \begin{minipage}[b]{\figwidth}
            \centering
            \raggedright (e)\\
            \includegraphics[width=\linewidth]{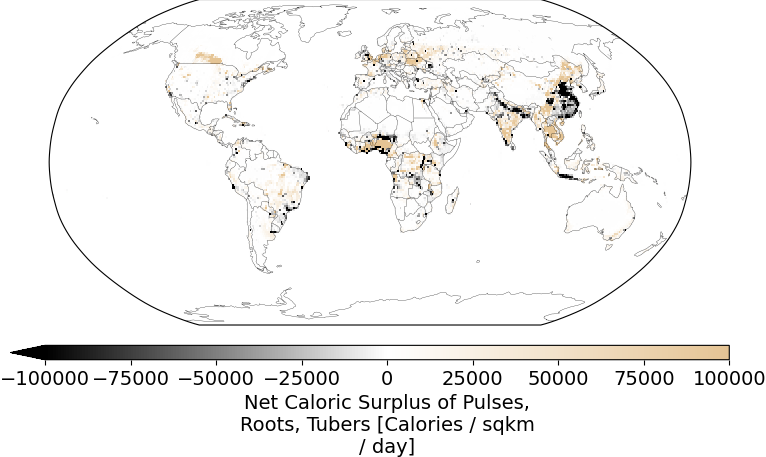}
        \end{minipage} &
        \begin{minipage}[b]{\figwidth}
            \centering
            \raggedright (f)\\
            \includegraphics[width=\linewidth]{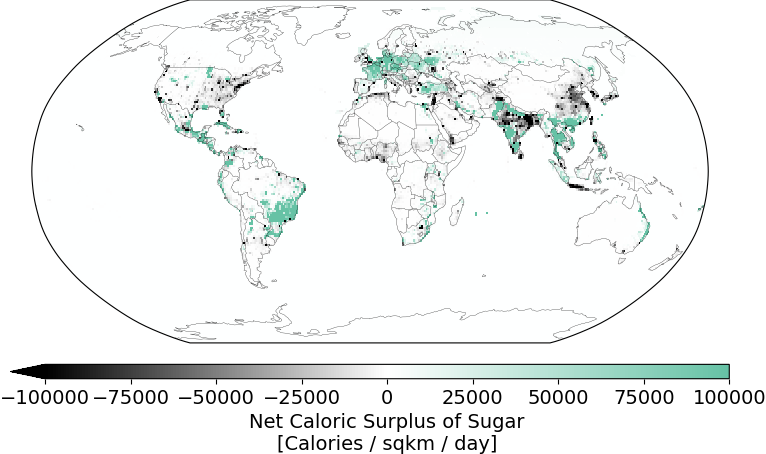}
        \end{minipage}
    \end{tabular}
    \caption{Net Caloric food balance maps. Each grid cell value stores the difference between caloric production of that group of foods and its corresponding consumption, where consumption is the sum of the feed and food supply values obtained by dasymetric mapping. Net producing grid cells are shown in color (corresponding to the associated color of that food group in figure 1. Net consuming grid cells are shown in gray scale. (a) livestock products as well as fish. (b) cereals. (c) fruits, vegetables, nuts. (d) oilcrops. (e) pulses, roots, tubers. (f) sugar. }\label{fig:nf_maps}
\end{figure}

Figure~\ref{fig:prod_cons} compares aggregated food production with the ultimate driver of the food system: human metabolism. Next we consider the spatial flow of food between grid cells with greater commodity and process resolution, by considering 6 food groups separately, and including the demand for livestock demand for food as well as human demand. The net flow balances of 6 groups of foods mapped in Figure~\ref{fig:nf_maps} account for each spatially explicit source and sink in our dataset by representing the sum of human food supply and animal feed subtracted from production for each food group. These maps show large differences in the spatial patterns of where particular food groups are produced relative to where they are consumed as food or feed. For example, most of India, China and the USA are characterized by local surpluses of cereal production but have large areas with deficits of fruit and vegetable production. The animal product net caloric balance map reveals that marine animal products (fish) are of minor caloric importance compared to terrestrial animal production in most of the world.

\subsection*{Metabolism and Food Supply}

\begin{figure}[htbp]
    \centering
    \begin{minipage}[b]{0.6\textwidth}
        \raggedright (a)\\
        \includegraphics[width=\linewidth,clip,trim={0 0 0 1.2cm}]{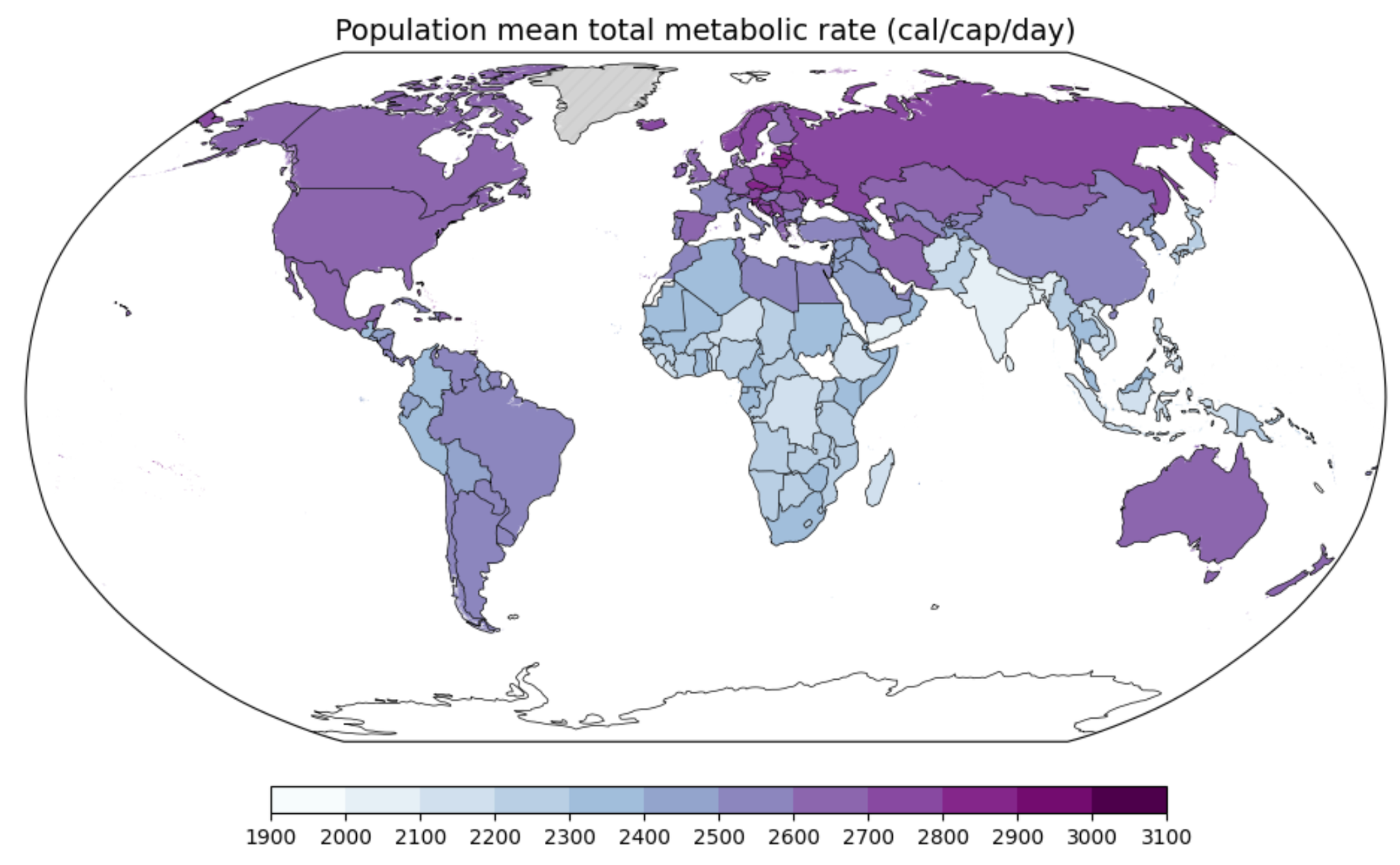}
    \end{minipage}
    \vspace{0.1cm}
    
    \begin{minipage}[b]{0.6\textwidth}
        \raggedright (b)\\
        \includegraphics[width=\linewidth]{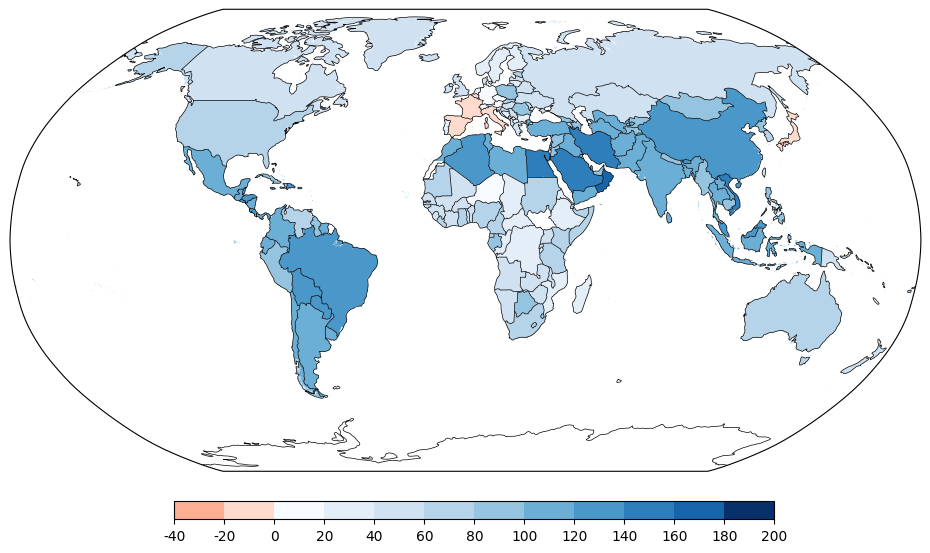}
    \end{minipage}
    \vspace{0.1cm}
    \caption{Average metabolic rates per capita showing spatial distribution and temporal changes. Units are kilocalries per day. (a) Population mean total metabolic rate per capita for the year 2015. (b) Changes in average per capita energy expenditure for each country available from 1990 to 2019.}
    \label{fig:ctry_met_map}
\end{figure}

Human metabolic rates vary substantially according to body size and activity level. We estimate that basal metabolic rates represent between 48\% and 69\% of the total metabolic rates for each country, with a mean Metabolic Equivalent of Task (MET, unitless ratio of total metabolic rate to basal metabolic rate, see Methods) value of 1.74 (1.44-2.04, 95\% confidence interval). This is consistent with prior estimates of population average MET between 1.42-2.00 \cite{pontzer_2021}. Figure \ref{fig:ctry_met_map} shows how average metabolic rates vary by country as well as the relative change between countries over the last three decades. Roughly half of the countries in the world have average total metabolic rates $<2500$ kcal/cap/day, most of which are in Africa, South Asia, and Southeast Asia, with 32 out of 34 low income countries falling below this threshold. Only four countries have decreased their average energy expenditure over this time, all of which are high income countries (Figure 4b and supplemental discussion S4).  While there is a tendency for lower income countries to have younger populations with smaller body sizes (due to age, mortality, and malnutrition), our results are consistent with prior work showing that total metabolic rates in lower income countries are not significantly different from higher income countries after controlling for body size and age \cite{dugasEnergyExpenditureAdults2011}.

\begin{figure}[htbp]
    \centering
    \begin{minipage}[b]{0.7\textwidth}
        \centering
        \raggedright (a)\\
        \includegraphics[width=\textwidth, trim={.5cm 0 1cm 0}, clip]{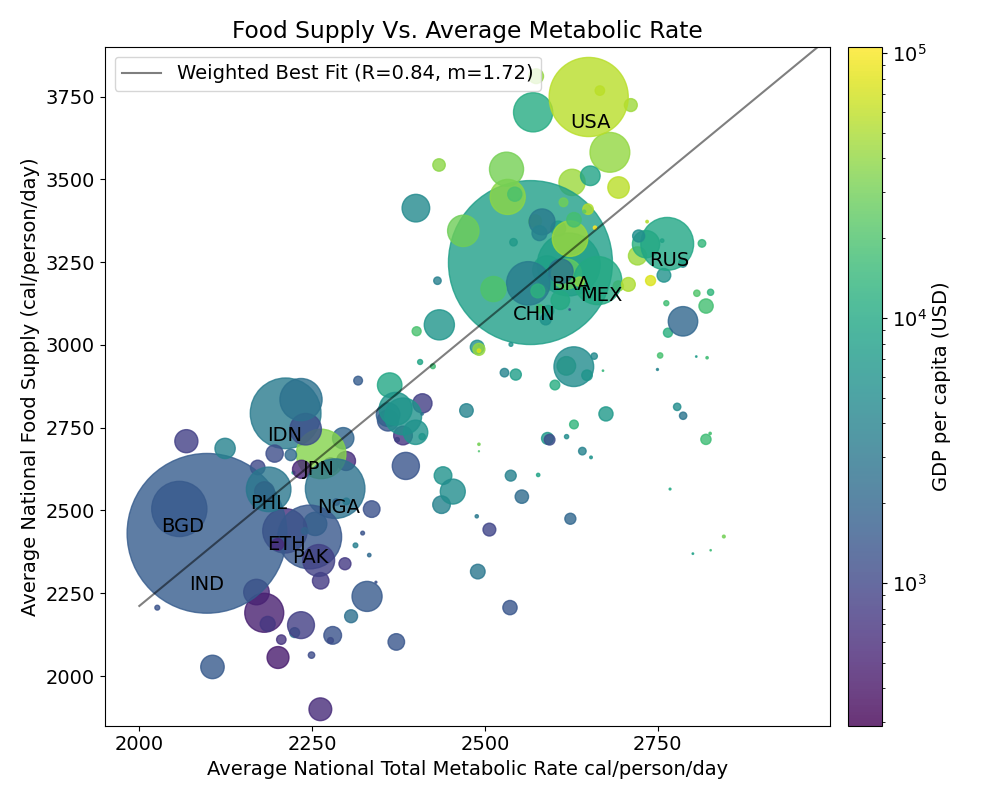}
        \label{fig:humanConsumption_a}
    \end{minipage}

    \begin{minipage}[b]{0.7\textwidth}
        \centering
        \raggedright (b)\\
        \includegraphics[width=\textwidth, trim={0cm 0 0cm 0}, clip]{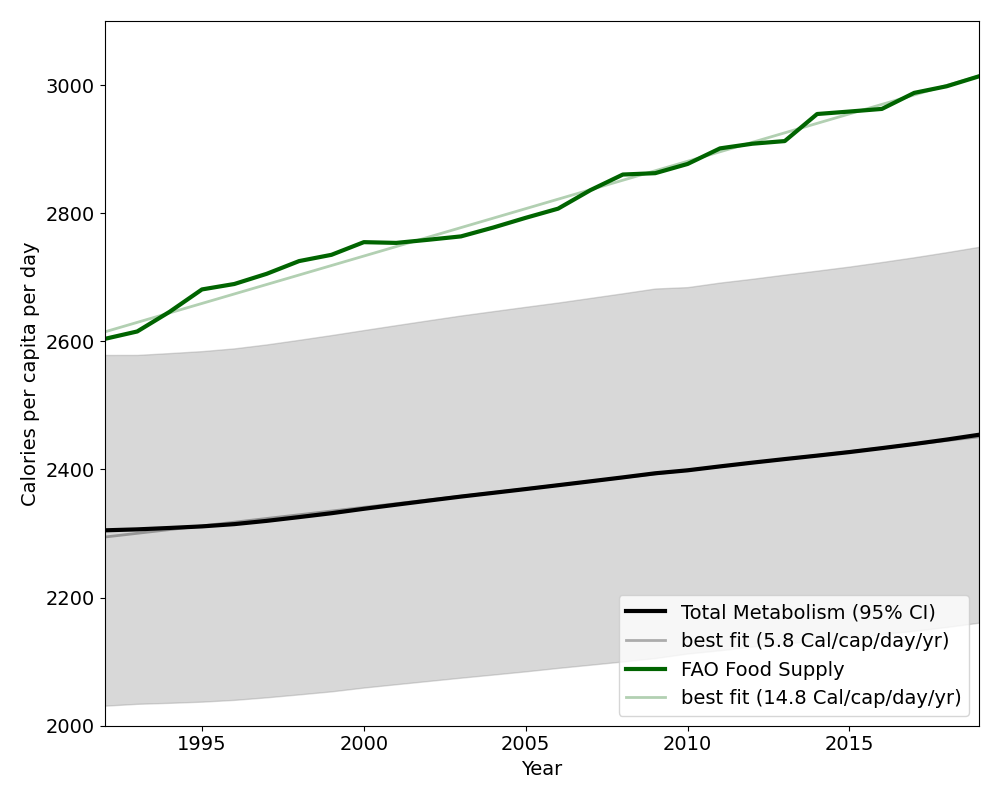}
        \label{fig:humanConsumption_b}
    \end{minipage}
    \caption{Relationship between food supply and metabolism. (a) Population weighted linear regression of national metabolic rates per capita vs food supply per capita for the year 2015. (b) Time series comparison of global food supply per capita and global metabolism per capita.}
    \label{fig:humanConsumption}
\end{figure}

Our estimates of global metabolic rates provide a benchmark for assessing efficiency in the food system. We build on previous work by estimating the fraction of the food supply that is not metabolized, either because it was lost or wasted (never consumed) or consumed and subsequently excreted or egested. We find a significant linear correlation between metabolic rates and food supply at the national level (Figure~\ref{fig:humanConsumption}a) ($p < .001$). The slope indicates that as the national metabolic rate per capita increases (due to greater body height and/or weight), national food supply per capita increases proportionally, at nearly twice the rate.

Our results show that a proportionally greater fraction of the available food energy is unmetabolized in countries with larger per capita food supplies (regression line with slope $>$1 in Figure 5a). In other words, countries with more food tend to waste (not metabolize) a larger fraction of it. While this relationship likely results from multiple factors, national gross domestic product (GDP) per capita is positively correlated with both food supply and metabolic rates, consistent with prior inferences of greater food waste in high-income settings \cite{Xue_2017}. An important underlying factor in this relative wealth-waste trend appears to be that people in wealthier countries tend to be larger, and therefore require more food to meet their metabolic needs. Interestingly, although GDP per capita is correlated with food supply, it weakly explains the residual variation with the best-fit line between metabolism and food supply, indicating that physical traits like height, weight, age, gender, and activity are perhaps more important (correlation between log GDP per capita and the unexplained residual, $r \approx 0.38$).

Global average metabolic rates increased over time, from 2283 kcal/person/day in 1990 to 2454 kcal/person/day in 2019, while the waste and excretion rate, calculated as (Food Supply - Metabolism) / Food Supply, nearly doubled from $\sim 10\%$ in 1990 to $\sim 20\%$ in 2019. This trend reflects the faster increase of global food supply compared to metabolism, consistent with an increase of consumer waste and inefficiency over the last three decades. We find that the food supply has increased linearly over time with an annual rate of about 15 calories per person per day while metabolic rates increased annually at 6 calories per person per day (Figure \ref{fig:humanConsumption}b). The average basal metabolic rate increased by 100 kilocalories over the last three decades due to the increase in the average size of humans, from 1295 kcal/cap/day in 1990 to 1395 kcal/cap/day in 2019.  
\subsection*{Metabolic Self-Sufficiency}
We calculate the potential metabolic self-sufficiency (MSS) for each food commodity $f$ in each region (grid cell or country) $i$ as the minimum of the region's metabolic demand and locally produced food for that food commodity. Metabolic demand accounts for total metabolism (both basal and active metabolism) as well as excreted calories, while locally produced food is estimated using a method similar to \cite{Cassidy_2013}. $MSS_{i,f}$ therefore represents the number of calories of metabolic demand that could be met ‘locally’ or domestically (within the same grid cell or country, respectively) by food-specific and region-specific production (Table \ref{tab:amss}). While holding national food commodity-specific diets constant, we find that approximately 51\% of the global population’s energetic requirements could be met by production in the same 1-degree grid cell. Relaxing dietary constraints to be food group-specific (following the groupings in Figure 1) rather than commodity specific could then sustain 62\% of people with local production. If metabolic demand were agnostic to food commodities altogether (sourced from any available calories including ‘Crops’, ‘Animal Products’, or ‘Processed Foods’), then 70\% of people could be sustained locally. At the country level, MSS is even higher, with 83 to 93\% of the global populations’ metabolic demand potentially satisfied without international trade.
\begin{table}[htbp]
    \centering
    \begin{tabular}{@{}l@{\hspace{.5em}}r@{\hspace{.5em}}r@{\hspace{.5em}}r@{\hspace{.5em}}r@{\hspace{.5em}}r@{\hspace{.5em}}r@{}}
        \toprule
        & \multicolumn{3}{c}{Population Sustained (\%)} & \multicolumn{3}{c}{Average MSS (kcal/cap/day)} \\
        \cmidrule(lr){2-4} \cmidrule(lr){5-7}
        Spatial Resolution $\downarrow$ & Commodity & Group & All & Commodity & Group & All \\
        \midrule
        Grid Cell & 51 & 62 & 70 & 1346 & 1636 & 1859 \\
        Country & 83 & 88 & 93 & 2224 & 2361 & 2489 \\
        \bottomrule
    \end{tabular}
    \caption{Metabolic Self Sufficiency ($MSS_{i,f}$) across spatial and commodity resolutions. The percentage of the population that is potentially food self sufficient with respect to a given spatial and commodity scale is calculated by summing the $MSS_{i,f}$ values for each food item within the same commodity classification $f$ in each region grid cell within the same spatial region $i$, and dividing by the total metabolic demand. This corresponds to the percentage of the global population that could sustain their metabolic demand if trade is closed within each spatial scale and if demand for calories of a given food can be met by any food in the same grouping at that commodity resolution. The $MSS_{i,f}$ values represent the total metabolic demand that could be met at each resolution, divided by world population. The food groups are the same as in Figure \ref{fig:energyFlowVornoi}.}
    \label{tab:amss}
\end{table}
\section*{Discussion}
We present a relatively straightforward but holistic framework to energetically balance the global food system and infer the spatial patterns of food energy production and consumption, including the ultimate rates of metabolism derived from demographic and activity data. By maintaining a high commodity resolution, we show distinct characteristics among the geographic distributions of different foods. We found a linear relationship between metabolic rate and food supply at the national scale, suggesting that food waste tends to grow proportionally with increases in per capita food supply. Additionally, our metabolism-based measure of waste, derived from a comparison of FAOSTAT data with entirely independent sociodemographic data \cite{Rodriguez_Martinez_2020,Phelps_2024}), showed that the unmetabolized fraction has increased over time. We further show the extent to which populations’ food-specific energy demands could be met at different spatial and commodity scales.

By estimating metabolism as a function of sociodemographic characteristics, rather than as a uniform distribution across the global population, our approach can contribute to future projections. The spread of obesity is expected to increase global consumption by the equivalent of 1 billion people \cite{Walpole_2012}, underscoring the importance of demographic consideration. Although food supply based on the FAOSTAT database is often used as a proxy for final food demand (as opposed to metabolic demand itself), we show that final metabolism has increased more slowly than the food supply globally, and that the metabolized fraction tends to decrease linearly with food supply between countries. Our study also provides a benchmark for country-specific caloric requirements. While previous studies have assumed a globally uniform caloric demand ranging from 1995 kcal/person/day \cite{Kummu_2012} to 2342 kcal/person/day \cite {alexander_2017} and 2700 kcal/person/day \cite{Cassidy_2013}, we estimate from socio-demographic data that national metabolic rates range between 2025 kcal/person/day (Timor Leste) and 2845 kcal/person/day (Bahamas) in 2015.

The food self-sufficiency of populations is of great interest, yet it does not have a unique definition, and prior work has shown that its quantitative assessment depends strongly on the chosen definition and methodology \cite{porkkaFoodInsufficiencyTrade2013, pradhanFoodSelfSufficiencyScales2014, Kinnunen_2020}. Our use of high commodity resolution and the complete set of food commodities tracked by the FAOSTAT database allows us to provide a new estimate of food self sufficiency distinct from prior works that considered only certain foods or food groups \cite{pradhanFoodSelfSufficiencyScales2014, Kinnunen_2020}. Notably, we estimate a higher food self-sufficiency than \cite{Kinnunen_2020}, who found that 11-28\% of the population's demand for 6 functional crop groups (accounting for about 47\% of globally traded calories) could be sustained within 100 km. We attribute at least part of this difference to the fact that \cite{Kinnunen_2020} minimized "food travel time," which could penalize trade within the same grid cell in favor of long distance trade (for example, grid cells with seaports) whereas we maximize the calories of metabolic demand that could hypothetically be met with current local production in the absence of trade. Further estimates were provided by \cite{pradhanFoodSelfSufficiencyScales2014} who found that in 2000, 1.9 billion people (or about 31\% of the global population) were food self-sufficient within 5-minute grid cells, while 4.4 billion people (or about 71\% of the global population) were food self-sufficient within their countries, and \cite{porkkaFoodInsufficiencyTrade2013} who found that 61\% of the global population lived in a country that met the entire country's self-sufficiency in 2005. The wide range across these studies indicates the importance of defining exactly what is intended by the term self-sufficiency. Our self-sufficiency metric, the minimum of locally produced calories and local energy demand, emphasizes the current capacity for calories being delivered to the food system within each grid cell.

Our approach facilitates easy updating, allowing for users to input new spatial demographic data. Future efforts to study food systems can easily adapt our framework by distributing more detailed national agri-food system data onto grid cells from different time periods, including country-specific caloric conversion factors, and adding subnational surrogate data for diets \cite{Kurtz_2020}. For example, we distribute some individual food commodities to crop groups in GAEZ+2015. Open source data could also be developed that includes locations of food processing and other supply chain hubs \cite{MacDonald_2023, Karakoc_Konar_Puma_Varshney_2023}, to assist in spatially downscaling production of processed foods and processing rates. Furthermore, spatial data on freshwater fish catch and aquaculture locations are lacking, despite comprising a large share of global fish supply \cite{Golden_2021}.

Our analysis also has implications for decision makers and policy. We have demonstrated that the subnational spatial patterns of metabolic self-sufficiency can provide an alternative perspective on food security and malnutrition, including both over- and under-consumption. The FAO reports that, despite there being $>$730 million undernourished people (calorie deficient), there are $>$880 million obese people (calorie excess), and this problem is only getting worse; the prevalence of undernourishment (caloric deficiency) increased from around 14\% to between 16\% and 19\% over the past decade in countries affected by conflict, climate extremes, or economic downturns, whereas countries not affected by a major driver of undernourishment maintained around 9\% \cite{fao2024stateOfFoodInsec}. This finding coincides with the trend we observed of increasing food waste (unmetabolized calories) in some countries (Figure \ref{fig:humanConsumption}b), suggesting a trend towards more waste alongside growing hunger. With increasing attention given to narratives of food sovereignty ~\cite{weilerFoodSovereigntyFood2015}, our framework provides a global quantitative approach to help contextualize discussions around local food systems. There have recently been promising efforts to balance and optimize land use demand for food production with species conservation \cite{hoangMappingPotentialConflicts2023} as well as with water demand and carbon storage \cite{bayerBenefitsTradeoffsOptimizing2023}. Such additional indicators could be added to our framework to navigate environmental tradeoffs.

Our analysis represents a step toward expanding the breadth of spatially explicit global food system analysis from crop photosynthesis to final metabolism with a relatively detailed tracking of specific food groups. We demonstrate how examining human metabolism can inform our understanding of food self-sufficiency and inefficiencies in the global food system by defining and accounting for waste as unmetabolized food supply. Further work on the global food system can continue to reveal emergent behavior of the billions of actors in the food system, which cannot be gleaned from the study of the sum of its local parts.

\section*{Methods}

Our method generally involves collecting and harmonizing national components of food production, transformation, and supply at the global scale, before downscaling these data geographically proportionally to spatial surrogates. We start by processing the Food Balance Sheet (FBS) \cite{FAO_FBS} data at the national scale, in order to maintain food energy conservation. Then in order to distribute food production, feed, and food supply to each grid cell, we obtain gridded surrogate data from a variety of sources, including global gridded crop production maps (GAEZ)\cite{GAEZ_2022}, the Global Livestock of the World Database\cite{FAO_Livestock_2018}, fish catch estimates\cite{Guiet_2024}, as well as population \cite{worldpop_2020}. We convert all food flow mass quantities into calories, using mass-energy conversion factors \cite{faoHandbook2001, Kastner_2011}\cite{MacDonald_2015}.

We compile a global 1-degree spatial resolution dataset that maintains a complete 95 food commodity resolution for grid-level variables including crop, animal and feed production, as well as food supply and metabolism. Human metabolism is estimated from size, demographic, and activity data, which is distributed among food commodities proportionally to each food's share of the local food supply, allowing comparison with other food-related flows and processes. We treat every food in the FAO Food Balance Sheet as a separate gridded data variable for each energy source (crop and animal production) and sink (livestock feed and metabolic demand). Table~\ref{tab:dataSources} outlines the FBS elements, their corresponding surrogate variables, and their data sources. For a given grid cell at latitude $i$ and longitude $j$ and food group $f$, the base data includes the following variables:
\begin{itemize}
   \item \textbf{Production} $P_{ij,f}$ : The total caloric energy produced (crops, livestock, fish).
   \item \textbf{Food Supply} $F_{ij,f}$ : The total caloric energy available for human consumption.
   \item \textbf{Animal Feed} $A_{ij,f}$ : The total caloric energy available for animal consumption.
   \item \textbf{Metabolism} $M_{ij,f}$ : The total caloric energy demand for human metabolism.
\end{itemize}

A complete list of all FAO Food Balance Sheet elements (corresponding to one of the primary data sources for each base variable) and their official definitions is provided in supplemental table S5. All units are given in terms of energy (1 kilocalorie = 4184 Joules of energy). We adopt a combination of these notations for the purposes of this study due to its interdisciplinary nature.

\begin{table}[htbp]
   \centering
   \begin{tabular}{|R{4cm}|R{2.5cm}|R{2.5cm}|R{4cm}|}
       \hline
       Derived grid variable & FBS Element source & Spatial surrogate variable & Surrogate variable source \\
       \hline \hline
       Human food supply & Food supply (kcal) & Population & WorldPop\cite{worldpop_2020} \\ \hline
       Animal feed & Feed & Livestock Units & GLW\cite{FAO_Livestock_2018}\\ \hline
       Crop production & Production & Crop production & GAEZ+2015\cite{GAEZ_2022} \\ \hline
       Animal product production & Production & Livestock units & GLW\cite{FAO_Livestock_2018}\\ \hline
       Marine food production & Production sum & Catch rate & BOATSv2\cite{Guiet_2024}\\
       \hline
   \end{tabular}
   \caption{Data sources and relationships. Column 1 gives types of inferred gridded data in the dataset, and column 3 gives the types of data used as a spatial surrogate for distribution. Column 2 gives the associated food balance sheet sheet source element and column 4 gives the surrogate data source for downscaling. Data from column 2 was mapped proportionally to the surrogate data in column 4 using the dasymetric approach in order to obtain local grid variables corresponding to column 1.}\label{tab:dataSources}
\end{table}

\subsection*{Food Balance Sheet data processing}

We first preprocess national food supply data, obtaining globally consistent annual production and consumption/utilization quantities for each FBS food commodity, in units of calories per year. The FBS methodology is predicated on the \emph{supply = utilization} identity (equation \ref{eq:fbs_supply_util})\cite{fbs_new_methodology, FAO_SUA}. The precise definitions of each element in equation \ref{eq:fbs_supply_util} is given by the FAOSTAT FBS methodology\cite{FAO_FBS} and are also provided in supplemental table S5.

\begin{align}
   &\text{Supply} = \text{Utilization}\ (for\ each\ country) \label{eq:fbs_supply_util}\\
   &\quad where \notag \\
   &\text{Supply} = Production + Imports - Exports + \Delta Stocks \notag \\
   &\quad and \notag\\
   &\text{Utilization} = Feed + Seed + Losses + Processing + Other\ uses\ \notag \\
   &\quad+ Tourist\ consumption + Residuals + Food \notag
\end{align}


\begin{gather}
   \text{Total } \textbf{Production } \text{(crop, animal, and processed)} = \text{Total } \textbf{Utilization} \notag \\
   \text{For each food }f, \sum_{\substack{i \in \text{countries} \\}} P_{i,f} = \sum_{\substack{i \in \text{countries}}} (F_{i,f} + A_{i,f} + Pr_{i,f} + OUL_{i,f})\label{eq:my_balance}
\end{gather}

Equation \ref{eq:my_balance} is analogous to the supply-utilization mass balance equation \ref{eq:fbs_supply_util}. Other Utilization and Loss ($OUL$) is a flex variable used to balance the rest of the data in equation \ref{eq:my_balance}, and represents the FBS columns $Seed$, $Losses$, $Other$ $Uses$, $Tourist$ $consumption$, and $Residuals$. We use the term production to refer to  edible animals or plant matter entering the global system (including crop growth, livestock growth, and processed food production), and utilization to refer to the edible plant and animal matter leaving the system (including food that is lost, wasted, used as biofuel, or else metabolized by humans or by animals). This choice is motivated by simplicity and data availability. Since alternative surrogate grid data exist (such as alternative cropland maps) and new datasets may become available, in principle, the other food balance sheet elements could be disaggregated and distributed onto the grid. Our approach builds off previous studies (e.g., \cite{Cassidy_2013}), and helps address problems with the SUA methodology, including sometimes heterogeneous national reporting methodologies\cite{FAO_SUA}.

\subsection*{Food energy conservation}

Given a region $i$ ($i$ can be a grid cell, a country, or any region of interest) and a food group $f$, the following equations must be true by the principle of conservation of energy. Equations \ref{eq:PU} and \ref{eq:IE} refer to global conservation equations for each food, summed over all regions globally, whereas equation \ref{eq:IPEU} is a family of equations for each each food and country individually (hence if there are x regions and y foods, we have 2x global constraints and xy local constraints).

\begin{equation}\label{eq:PU}
\sum_i P_{i,f} = \sum_i U_{i,f}
\end{equation}
\begin{equation}\label{eq:IE}
   \sum_i I_{i,f} =  \sum_i E_{i,f}
\end{equation}
\begin{equation}\label{eq:IPEU}
P_{i,f} + I_{i,f} = U_{i,f} + E_{i,f}
\end{equation}

\noindent where $I_{i,f}$, $P_{i,f}$, $E_{i,f},$ and $U_{i,f}$ denote a region's imports, production, exports, and utilization, respectively. The utilization term $U_{i,f}$ is defined as $U_{i,f} = A_{i,f} + Pr_{i,f} + Ol_{i,f} + F_{i,f}$ (and hence equation \ref{eq:PU} is a generalization of equation \ref{eq:my_balance} to arbitrary regions i). Equation~\ref{eq:IE} follows from the notion that the trade system is closed; all imported food must have been exported from elsewhere. Equation~\ref{eq:IPEU} describes how food can enter a region only through production or imports, and it can leave only through utilization (including both food consumption and non-food utilization, loss, and waste) or exports.

Notably, equations~\ref{eq:PU},~\ref{eq:IE}, and~\ref{eq:IPEU} are not dependent on food system nodes being countries or grid cells. The system of equations describes the conservation of mass or energy of any closed food system network at any scale. This means that the equations hold whether the index set labels nations, grid cells, or any other food system network, as long as food only enters the system through $P_i$ and only leaves through $U_i$, and trade is closed between nodes in the system.

In principle, the total import quantity should be equal to the total export quantity, since all exported food must go somewhere (except for relatively small amounts of food that are lost in transit). However, trade data asymmetries are known to exist due to re-exports, insurance policies, tariffs on imports, misreporting and under-reporting, among other factors \cite{MacDonald_2015, Kastner_2011, Markhonko_2014, Kellenberg_Levinson_2019}. In order that equation~\ref{eq:fbs_supply_util} is satisfied, the FBS therefore must have discrepancies in other fields as well in order to make the equation balanced with imbalanced imports and exports. In the FBS, global total import quantity exceeds total export quantity (by about 20 Megatons of food in 2015), which places approximately 14\% of global food trade (4\% of the human food supply) in apparent violation of equation~\ref{eq:IE}. Compounding this, in order to satisfy equation \ref{eq:IPEU} at the national scale, the FBS reports residual differences when the equation does not balance.

To address these inconsistencies, we adjust the reported values by solving a constrained quadratic programming problem that minimizes the relative root mean squared error (RMSE) between the adjusted and original values, while enforcing the conservation constraints in equations~\ref{eq:PU},~\ref{eq:IE}, and~\ref{eq:IPEU} for each food and country. Specifically, for each food commodity, and for each country $i$, we define $P_i$ as the adjusted production, $E_i$ as the adjusted exports, $U_i$ as the adjusted utilization, $I_i$ as the adjusted imports, and $P_i^{0}, E_i^{0}, U_i^{0}, I_i^{0}$ as the original reported values.
We solve the objective function:
\begin{align*}
   \min_{\{P_i, E_i, U_i, I_i\}} \quad & \sum_i \left(
       \frac{(P_i - P_i^{0})^2}{(P_i^{0} + \epsilon)^2}
       + \frac{(E_i - E_i^{0})^2}{(E_i^{0} + \epsilon)^2}
       + \frac{(U_i - U_i^{0})^2}{(U_i^{0} + \epsilon)^2}
       + 2\frac{(I_i - I_i^{0})^2}{(I_i^{0} + \epsilon)^2}
   \right) \\
   \text{subject to the constraints:} \quad
   & P_i + I_i = E_i + U_i \quad \forall i \\
   & \sum_i P_i = \sum_i U_i \\
   & \sum_i E_i = \sum_i I_i \\
   & P_i \geq 0,\; E_i \geq 0,\; U_i \geq 0,\; I_i \geq 0 \quad \forall i
\end{align*}

where $\epsilon$ is a small constant to avoid division by zero. We include a coefficient of 2 in front of the import term in the objective function to penalize relative errors in imports more heavily. This reflects evidence that countries are more likely to accurately track and report their own imports \cite{Kellenberg_Levinson_2019}, making import data generally more reliable than export data. By increasing the cost of deviating from reported import values, the optimization prioritizes matching imports, and adjusts exports accordingly to satisfy equation~\ref{eq:IE}. We also proportionally scale all production values by 2\% to ensure that the total production matches total utilization, as required by equation~\ref{eq:PU}. Details of the optimization are provided in supplemental discussion S2.1.

Next, we convert production, feed, and processing into units of kilocalories (kcal). For each food, the energy density is obtained using caloric conversion factors ~\cite{faoHandbook2001, MacDonald_2015}. For a small number of food commodities where conversion factors were unavailable, we infer the energy density as the fraction of the FBS 'Food supply (kcal)' divided by the 'Domestic Supply Quantity' for each country and year. We then use this energy density to convert each element in the food balance sheet (production, processing, imports, exports,  food, feed, etc.) to units of kilocalories/year. While more than half of crop production energy flows to the food supply either directly, via animal product conversion, or via food processing, there are some notable exceptions. For example, soybean processing for livestock feed produces oil for human consumption, entailing links between processed food production and feed as well as between animal products and processing (shown as "by-products and co-products" in Figure \ref{fig:energyFlowVornoi}). The processing of soybeans involves crushing the beans into roughly 80\% cake, commonly used as animal feed, and 20\% oil for human food \cite{agriculturalConversionFactors2061,Cassidy_2013,FAO_FBS}. Dried distillers grains (DDGs) are an example of a co-product of ethanol production, and play an increasing role in the flow of calories in the North American food system in particular\cite{olsonDriedDistillersGrains2019}.

Our choice to use calories as the unit of food flows was in part motivated by recent research interest in the network of food flows around the world \cite{Schreiber_2021}, and its impact on sustainability, biodiversity, and nutrition \cite{kastnerGlobalAgriculturalTrade2021}. There have been efforts to study mass food flows between subnational scales ~\cite{konarScalingPropertiesFood2018, linFoodFlowsCounties2019, dongFrameworkQuantifyMass2022, karakoc_wang_konar_2022, panditSpatiallyDetailedAgricultural2023}, some balancing the trade-off between efficiency and resilience of food supply chains \cite{karakocComplexNetworkFramework2021}. Mass and dollars are the conventional units for studying food flows in this context, largely due to the convenience of reporting and studying production and trade. Kilocalories are, however, a more meaningful unit around food supply and metabolism. While some have shown that global food trade has the potential to help reduce food insecurity \cite{Smith_2021, DOdorico_2014}, others have argued that the international fish trade plays a role in further diverting nutrients from nutrient-insecure regions\cite{nashTradeForeignFishing2022}. Having studied the food flows at first order (i.e. production and consumption in individual grid cells and regions), our data and pipeline provide a foundation to increase both the detail and the completeness for future second order (i.e. trade between pairs of locations) food flows and food security assessments in an effort to balance these concerns.

\subsection*{Metabolism of the human population}

We approximate human metabolism globally using a simplified version of the Mifflin-St Jeor Equation \cite{Mifflin_St_Jeor_1990} (equation~\ref{eq:bmr}), averaged between males and females to estimate the average basal metabolic rate of individuals from each country. The Mifflin-St Jeor equation outperforms the other three most commonly used equations in clinical practice when compared to measured resting metabolic rate data\cite{frankenfieldComparisonPredictiveEquations2005}.

\begin{equation} \label{eq:bmr}
   BMR_i = 10 W_i + 6.25 H_i - 5 A_i - 161 G_i + 5 (1 - G_i)
\end{equation}

where $BMR_i$ is Basal Metabolic Rate in kilocalories/day, $W_i$ is average weight in kg, $H_i$ the average height in cm, $A_i$ is the average age in years for the $i$th country, and $G_i$ is the number of women divided by the number of men in the $i$th country. Height and weight data are obtained from Lancet pooled research report estimates\cite{Rodriguez_Martinez_2020,Phelps_2024}. Population age and gender demographics were obtained from UN estimates\cite{UNPopProspects}. Basal metabolic rate is then distributed proportionally to population density using the dasymetric approach.

We obtain an estimate of total metabolic energy expenditure for each country by multiplying each country's basal metabolic rate by the MET score associated with the average activity level of the population. 1 MET is the metabolic equivalent of a task, defined as the ratio of energy expenditure of a task to basal metabolic rate.

\begin{equation} \label{eq:tmr}
   M_i = BMR_i * \sum_{j} MET_{j} * A_{i,j}
\end{equation}

556 MET activities from \cite{herrmann2024CompendiumPhysical2024} were mapped to the time use activities from \cite{Fajzel_2023} using a concordance matrix (supplemental table S1). Each time use activity had several MET activities associated with it, with some high-value outliers present. We took the median MET for time use activity $j$ as the representative value $MET_{j}$.

\subsection*{Metabolism of animals}

We calculate the total Livestock Units (LSU) of livestock in the world at the grid level first by multiplying livestock counts by region-specific LSU values (supplemental table S3). We then compute the total metabolism of livestock by multiplying the total LSU by the average metabolism of one LSU, which has been estimated to require approximately 75 MJ/day of metabolizable energy (ME) \cite{mokolobateExplainingPrincipleLarge2017}.

\begin{align*}
   \text{Total metabolism of livestock} &= \sum_{i} \text{LSU}_i \times \text{kcal/day/LSU} \\
   &= 1.8*10^{15} LSU \frac{75 MJ}{LSU day} \frac{1 kcal}{4.184*10^{-3} MJ} \frac{population}{7.4*10^9} \\
   &= 4360 \text{ kcal/capita/day}
\end{align*}

In order to roughly approximate the metabolism of the global marine fish catch, we assume that fish metabolism is roughly 4.25 J/mg wet mass\cite{breyBodyCompositionAquatic2010}, that the fish catch biomass is 137 Mt/year \cite{FAO_FBS}, that fish metabolize 5 times their biomass annually\cite{bianchiEstimatingGlobalBiomass2021}, and that the average age of catch is about 2 years (a rough order of magnitude assumption, given that the true value is unknown)\cite{ReviewLengthbasedApproaches}. We estimate the total metabolism of fish as:
\begin{align*}
   \text{Total metabolism of fish} &= \text{Wet Biomass} * \text{Metabolism per mass} \\
   &\sim \frac{2 fish}{catch} \frac{5 * 137 Mt}{year} \frac{1 year}{365 days} \frac{10^{15} mg}{Mt} \frac{4.25 J}{mg} \frac{1 kcal}{4184 J} \frac{population}{7.4*10^9} \\
   &\sim 500 \text{ kcal/capita/day}
\end{align*}





\subsection*{Spatial data allocation}

We next populate a 1-degree resolution global grid (approximately 10,000 km$^2$ per grid cell at the equator), by distributing FBS data into each grid cell proportionally to local variables (surrogate variables). Our choice of a relatively coarse 1-degree resolution is intended to provide a unified global perspective on the food system that captures subnational patterns without attempting to quantify local (e.g., city-level) flows. This resolution provides both the computational feasibility for polynomial time global grid analyses (such as analysis of the network of food flow between grid cells), as well as the granularity to visualize and compare the roles of population centers with large agricultural regions and hybrid areas, without being overwhelmed by fine-scale detail. We obtain gridded maps of crop production for the year 2015 from GAEZ+2015\cite{GAEZ_2022}, which provides local surrogate data for production of crop foods in the FBS. Livestock counts are obtained from Gridded Livestock of the World\cite{FAO_Livestock_2018}, which form the basis for LSU surrogates for both feed consumption as well as animal product food production. Estimates of wild marine fish capture rates from 2015 are obtained from the data-constrained BoatSv2 model\cite{Guiet_2024}, which provide surrogate data for marine fishery production (see supplementary discussion S3.3 for alternative catch surrogates). The global human population map is obtained from WorldPop, providing the basis for food demand and metabolism per grid cell \cite{worldpop_2020}.

We then downscale country-level food production, feed, and food supply data from the 2015 FAO FBS proportionally to their associated grid-level surrogate variables. This procedure, known as dasymetric mapping, has a long history in cartography and geography \cite{Petrov_2012}. We chose dasymetric mapping as our approach for spatial allocation and downscaling, as it is a simple and non-trivial downscaling algorithm: regional data is distributed proportionally to a single grid-level surrogate. We seek to enable study of the relationship across scales, and hence need to rely on both local information and regional data. By distributing each regional quantity proportionally to a single surrogate quantity, we ensure easy reproducibility and interpretability of results. We formalize dasymetric mapping and distribute the data according to the dasymetric equation (equation~\ref{eq:das_eq}, described in detail in~\cite{SESAME}). $X_C$ represents national scale data for country $C$, and $x_{ij}$ is the corresponding distributed grid data at latitude $i$ and longitude $j$. $C_{ij}$ represents the area of country $C$'s jurisdiction over grid cell $ij$, and  $s_{ij}$ represents surrogate data in that grid cell.

\begin{equation} \label{eq:das_eq}
   x_{ij} = X_C * \frac{s_{ij}}{\sum_{ij \in C} C_{ij} s_{ij}}
\end{equation}

We chose 2015 as a base year since it is the middle of the time-span represented by our input dataset (livestock counts are from 2010 whereas crop production data is from 2015). Table~\ref{tab:dataSources} summarizes the data sources and associated surrogate variables. The supplemental information provides details about specific food items and groups and each item’s proportion of the total food supply (table S4), as well as their correspondences with crop and livestock production surrogates (tables S6, S7, and S8). The foods that are distributed to the grid using a surrogate variable account for about 97\% of the food supply and 99\% of production reported by FAOSTAT for 2015. The remaining 1\% of production corresponds to food commodities and types of utilization that lacked surrogate data, most importantly fresh water fish and aquaculture. Other FBS food commodities without a suitable surrogate were 'Beverages, Fermented', 'Beverages, Alcoholic', 'Pepper',  and 'Sweeteners, Other'.

We compute the net caloric production in each grid cell for each food group as the difference between the production and the supply of that food. For some countries which have production values in the FBS, but no available corresponding surrogate variable in the country, production is distributed uniformly among grid cells.

\subsection*{Metabolic self sufficiency}

We define metabolic self-sufficiency in equation \ref{eq:mss}, by comparing rescaled production to reflect the subset of production that will eventually enter the food supply (equation \ref{eq:available_prod}) with metabolic demand (including 5\% excretion).

\begin{equation} \label{eq:mss}
   MSS_{i,f} = \min(AP_{i,f}, M_{i,f})
\end{equation}

The available production $AP_{i,f}$ is calculated as:

\begin{equation} \label{eq:available_prod}
   AP_{i,f} = (P_{i,f} - E_{i,f})\frac{F_{i,f}}{P_{i,f} + I_{i,f} - E_{i,f}} + E_{i,f} \sum_{j \neq i}\frac{F_{j,f}}{P_{j,f} + I_{j,f} - E_{j,f}}\frac{I_{j,f}}{\sum_{k \neq j} I_{k,f}}
\end{equation}

A similar approach was proposed by \cite{Cassidy_2013} and was used by \cite{Kinnunen_2020}. The term $\frac{I_{j,f}}{\sum_{k \neq j} I_{k,f}}$ rescales the exported portion of production from country $i$ based on the domestically available food supply $P_{j,f} + I_{j,f} - E_{j,f}$ in each importing country $j$. The metabolic demand $M_{i,f}$ is calculated by multiplying each grid cell's population by their national per-capita total metabolic rates from equation \ref{eq:tmr} by a factor of 5\%, based on existing estimates for energy rates of excretion \cite{basoloEffectsUnderfeedingOral2020, heymsfieldEnergyMalabsorptionMeasurement1981}.

\subsection*{Uncertainty}

Our global food system assessment compiles data from multiple sources and across phases of the food system, hence there are many possible sources of uncertainty and inaccuracy. We rely on calorie density conversion factors, assumed to be uniform for each food group, but which in reality vary geographically and between processing stages or food preparation. For instance, the caloric density of a crop at harvest on the farm might be different than further downstream in the food supply chain. The spatial heterogeneity of food energy densities has been addressed in some datasets \cite{smithGlobalExpandedNutrient2016}, which can be incorporated in future updates to the open source data pipeline. On the consumption side, while demographic average weights and heights, as well as food supplies, were distributed from the country level according to gridded populations, these distributions are not necessarily uniform in reality.

The global scale of the analysis also entails sources of uncertainty and accuracy. Throughout this study we have attempted to use the best available resources and methods to maintain global parsimony and interpretability while minimizing the number of assumptions made. In an effort to mitigate uncertainty at the global scale, we use a top-down approach to downscale harmonized Food Balance Sheet data. However, the global sums result from a conglomeration of a variety of nationally reported data sources \cite{FAO_FBS}, with varying levels of fidelity. Finally, there is uncertainty in the relationship between the surrogate data and the spatial distribution of the national scale data. As described by \cite{SESAME}, the assumption that national data is spatially distributed proportionally to a single surrogate data source is a simplifying assumption with complex underlying uncertainties.

\bibliographystyle{plainnat}
\bibliography{foodGrid}

\end{document}